\documentclass[9pt,twocolumn,twoside]{osajnl}
\journal{ol} % Choose journal (ao, aop, josaa, josab, ol, pr)

\setboolean{shortarticle}{true}
\title{Correlations in scattered perfect optical vortices}

\author[1,**]{Patnala Vanitha}
\author[2,3]{Nijil Lal}
\author[2,3]{Anju Rani}
\author[1,*]{Salla Gangi Reddy}
\author[2]{R. P. Singh}

\affil[1]{Department of Physics, SRM University-AP, Amaravati, India - 522508.}
\affil[2]{Physical Research laboratory, Navarangpura, Ahmedabad, India-380 009.}
\affil[3]{Indian Institute of Technology, Gandhi Nagar, India - 382355.}

\affil[*]{Corresponding author: gangireddy.s@srmap.edu.in}
\affil[**]{vanitha\_patnala@srmap.edu.in}

\begin{abstract}
We have studied correlations in the speckle patterns generated by the scattering of perfect optical vortex (POV) beams and used them for producing a new-class of coherence functions, namely Bessel coherence functions. Higher (zeroth) order Bessel coherence functions have been realized in cross (auto)-correlation between the speckle patterns generated by the scattering of perfect vortex beams of different orders. We have also studied the propagation of produced Bessel coherence functions and characterized their divergence with respect to the radius of their first ring for different orders $m$=0--4. We observed that the divergence varies linearly with the order of the coherence function. We provide the exact analytical expression for the auto-correlation as well as cross-correlation functions for speckle patterns. Our experimental results are in good agreement with the analytical results. 
\end{abstract}

\setboolean{displaycopyright}{true}

\begin{document}

\maketitle

Optical vortices are well known for phase singularity at their centre with screw dislocations \cite{Nye}.  These beams have helical wave fronts with an azimuthal phase of $e^{im\theta}$ and carry an orbital angular momentum of $m\hbar$  per photon, where $m$ is the topological charge (TC) or order of the vortex which can be defined as number of helical wave fronts completed in one wave length. The vortex beams have found many applications in science and technology \cite{applications, torner} such as optical communication using the infinite dimensional basis states. Optical vortices have dark core at the center whose diameter depends on the order. In order to control the intensity distribution of vortices, a new type of vortices have been introduced, named as perfect optical vortices (POV) \cite{ostro}. These beams have a topological charge independent intensity distribution with controllable ring size and ring width. Many techniques have been proposed and verified to generate the POV beams and utilized in various applications \cite{pravin, carvajal, parmoon, chen, sgreddy}.

The correlation in light beams is becoming one of the promising fields of study since it plays an important role in many of the applications such as communication and cryptography \cite{mandel, gbur, nischal}. The correlation functions that have singularities are named with a new term “Coherence vortices”  \cite{random, visser, wang, alves}. The correlations in the partially coherent beams can be described with cross-spectral density (CSD) function, and have advantages over that of a coherent beam in some applications such as free space optical communication, remote sensing and optical imaging \cite{paterson, david, cai, kermi, peng}. Recently, these coherence vortices are realized in the intensity correlation of two speckle patterns obtained by the scattering of coherent vortex beams of different orders \cite{alves, goodman}. 

In this study, we investigate theoretical and experimental investigation on the scattering of POV beams for generating a special class of coherence vortex, namely, Bessel coherence function. These functions have been generated with the help of intensity correlation between two speckle patterns corresponding to POV beams of different orders. We have also provided the exact analytical expression for the correlation function which are in good agreement with the experimental results.

We start with the field distribution of a POV beam, described by the thin annular ring of order $m$, given by \cite{ostro}
\begin{equation}
E(\rho,\theta)= \delta\left(\rho-\rho_0\right)\ e^{im\theta}
\label{field}
\end{equation}
where $\rho_0$ is the radius of the POV beam and $\delta$ represents the Dirac delta function. The scattering of POV beams through a ground glass plate (GGP) can be described with random phase function $e^{i\Phi}$, where $\Phi$ varies randomly from 0-2$\pi$. Now, the field distribution of speckles after scattering through the GGP is given by \cite{goodman}
\begin{equation}
U(\rho,\theta)= \delta\left(\rho-\rho_0\right)\ e^{im\theta} *e^{i\Phi}
\label{Sfield}
\end{equation}
In this study, we are interested in finding the  mutual coherence function between the two scattered POV fields. $\Gamma\left( .\right)=U_1(r_1,\varphi_1,z)U_2^*(r_2,\varphi_2,z)$  has been evaluated by using the Fresnel diffraction integral in cylindrical coordinates, which can be represented mathematically as \cite{acevedo}

\begin{eqnarray}
\Gamma_{12}\left( \Delta r\right)=\frac{e^{\frac{ik}{2z}\left( r_1^2-r_2^2\right) } }{\lambda^2 z^2}\int\int  U_{m_{1}}(\rho,\theta)U_{m_{2}}^*(\rho,\theta)  \nonumber  \\
e^{-\frac{ik}{z}\left( \rho\Delta rcos(\varphi_s-\theta)
\right)} \rho d\rho d\theta
\label{Correlation}
\end{eqnarray} 
  
where
\begin{eqnarray}
\Delta rcos(\varphi_s-\theta)&=&\left[\left( r_1 cos\left( \varphi_1\right) -r_2 cos\left( \varphi_2\right)\right)cos\theta\right]             \nonumber  \\
& + &  \left[\left( r_1 sin\left( \varphi_1\right) -r_2 sin\left( \varphi_2\right)\right)sin\theta\right]
\end{eqnarray} 
and $\Delta r^2=r_1^2+r_2^2-2r_1r_2 cos\left(\varphi_2-\varphi_1\right) $. 

The cross-correlation function of two speckle patterns corresponding to two different orders ($m_1 \&  m_2$) of POV beams is given by the Bessel function of order $m = m_2-m_1$. The complete mathematical expression is given by 
\begin{eqnarray}
\Gamma_{12}\left( \Delta r\right)&=&\frac{2\pi \left( -1\right) ^{{m_{2}}-{m_{1}}}i^{{m_{2}}-{m_{1}}}e^{\frac{ik}{2z}\left( r_1^2-r_2^2\right)} }{\lambda^2 z^2}e^{i\left( {m_{1}}-{m_{2}}\right)\varphi_s}
\nonumber  \\ 
& & \int ^{\infty} _{0}\delta\left(\rho-\rho_0 \right) J_{\left({{m_{2}}-{m_{1}}}\right)}\left( \frac{k\rho}{z}\Delta r\right)   \rho d\rho
\label{CCorrelation}
\end{eqnarray}

Using Eq. \ref{Sfield} in the above equation with the help of Anger-Jacobi Identity and the integral properties of Dirac-delta function \cite{ryzhik}, we get

\begin{eqnarray}
\Gamma_{12}\left( \Delta r\right)&=&\frac{2\pi\rho_0 \left(-i\right)^{{m_{2}}-{m_{1}}}e^{\frac{ik}{2z}\left( r_1^2-r_2^2\right) } }{\lambda^2 z^2 } \nonumber \\
& & e^{i\left( {m_{1}}-{m_{2}}\right)\varphi_s}  J_{\left({{m_{2}}-{m_{1}}}\right)}\left( \frac{k\rho_0}{z}\Delta r\right).  
\label{output}
\end{eqnarray}
From the above equation, it is clear that the cross-correlation function of two speckle patterns is well described by the Bessel function of order $m=m_2-m_1$.

The corresponding complex coherence function of the scattered field is given by  
\begin{equation}
C( \Delta r)= (-i) ^{{m_{2}}-{m_{1}}} e^{i( {m_{1}}-{m_{2}})\varphi_s} J_{{{m_{2}}-{m_{1}}}}\left( \frac{k\rho_0}{z}\Delta r\right) 
\end{equation}
Normalized intensity distribution of the coherence function can be evaluated in terms of time averaged intensity $I_0$ as
\begin{eqnarray}
I\left( \Delta r\right)&=&I_0^2 \left(1+\vert\mu_u\left( \Delta r\right)\vert ^{2} \right)  \nonumber \\
&=&I_0^2 \left(1+J_{\left( {{m_{2}}-{m_{1}}}\right) }^2\left( \frac{k\rho_0}{z}\Delta r\right)\right)
\label{output2}
\end{eqnarray}

If two speckle patterns are corresponding to the same order, the cross-correlation function will be converted as auto-correlation function which can be obtained by keeping $m_1=m_2$ in the above equation. We get the auto-correlation function as 

\begin{equation}
I\left( \Delta r\right)=I_0^2 \left(1+ J_0^2\left( \frac{k\rho_0}{z}\Delta r\right)\right) 
\label{output3}
\end{equation}

It is clear from the above analysis that the auto-correlation functions can be described with Bessel functions of order zero and cross-correlation functions can be described with Bessel-functions of non-zero orders ($m=m_2-m_1$). 

\begin{center}
   \begin{figure}[h]
   \includegraphics[width=8.0cm]{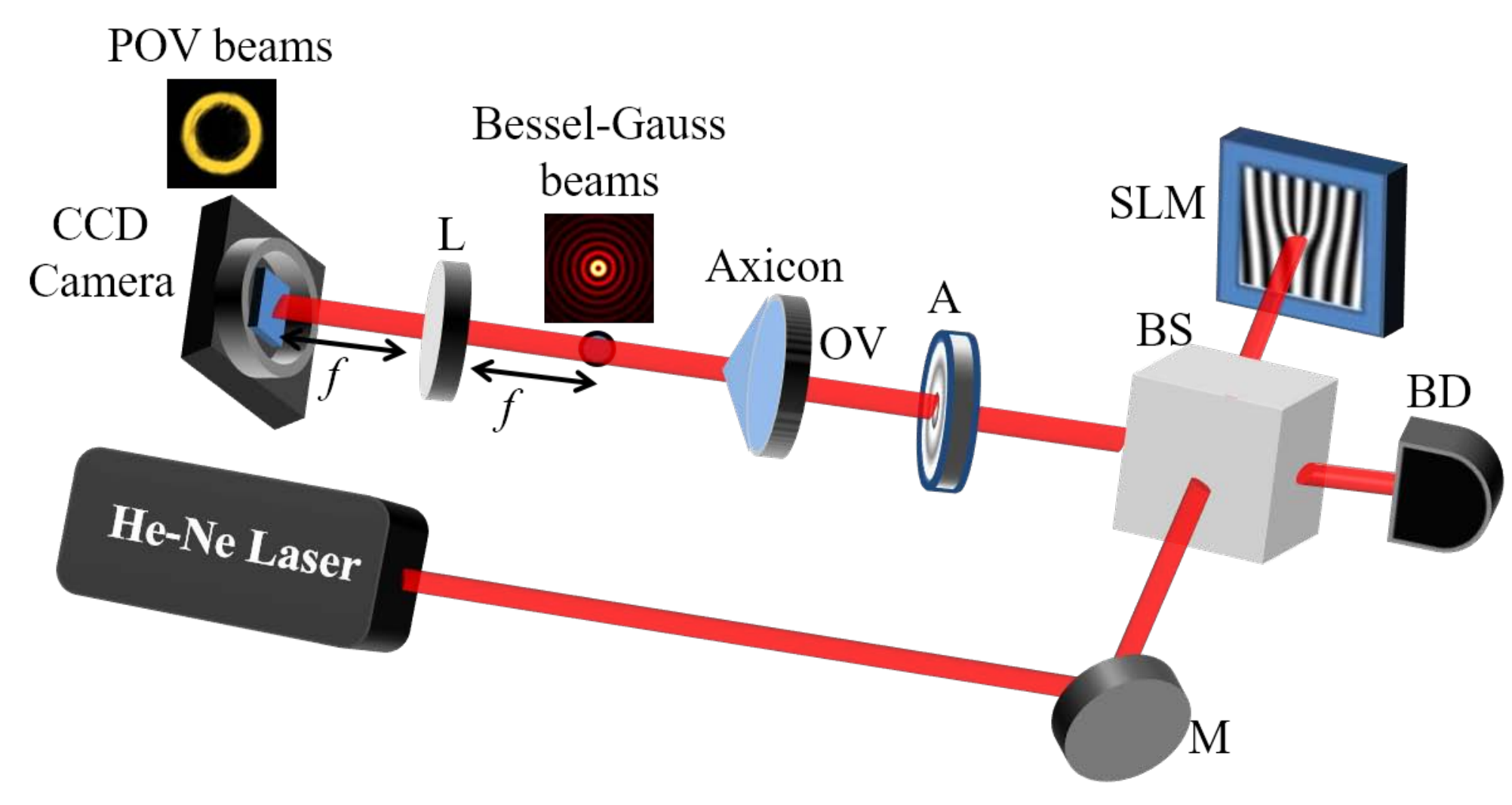}
    \caption{\textit{(Colour online) Experimental setup for the generation of perfect optical vortex beams using a spatial light modulator and an axicon. A-Aperture, M-Mirror, SLM-Spatial Light Modulator, BD-Beam Dumper, \textit{f}-Focal length}}
    \label{fig:expt}
   \end{figure}
   \end{center}

 Experimental setup for the generation of POV beams is shown in Fig. \ref{fig:expt}. Optical vortex beams have been generated using computer generated hologram displayed on a spatial light modulator (Holoeye LCR-2500) with the help of a He-Ne laser beam of wavelength 632.8 nm and power 5 mW. The desired vortex beam selected with an aperture (A) is passed through an axicon, with an apex angle of $178^{0}$, in order to convert the vortex 
beam into Bessel-Gauss beams. The axicon is placed at a distance of 50 cm from SLM. The Bessel-Gauss beams have been generated at a distance of 12.5 cm from the axicon. This has been considered as the plane of generation for BG beams. A lens (L) of focal length 30 cm has been placed at the focal plane, which will provide the Fourier Transform of the BG beams, for creating the POV beams. A CCD camera (FLIR, pixel size 3.45$\mu$m) is placed at the Fourier plane for recording the intensity distribution of POV beams.

\begin{center}
   \begin{figure}[htb]
   \includegraphics[width=8.0cm]{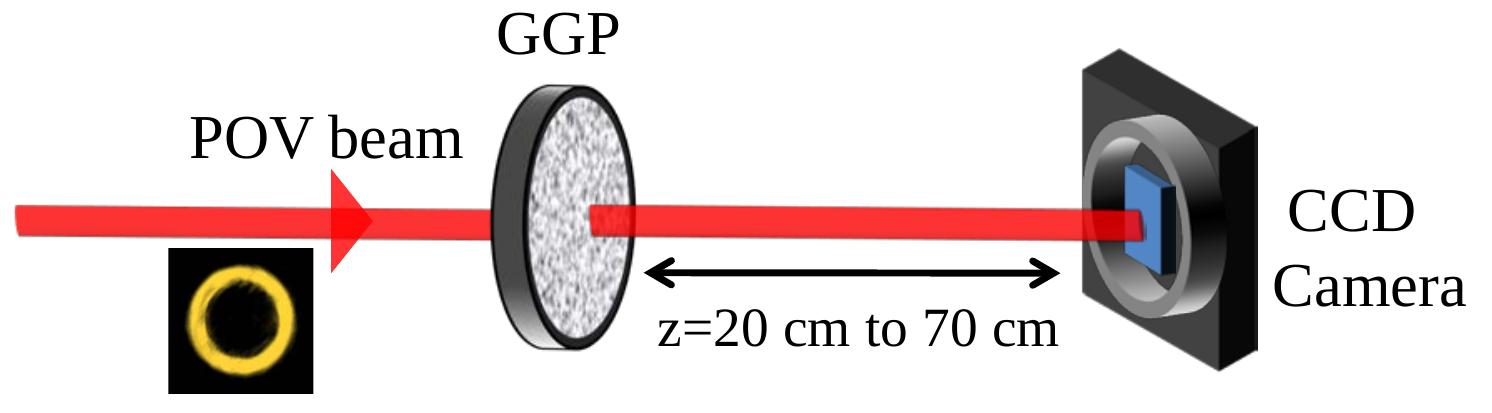}
    \caption{\textit{(Colour online) Experimental set up for recording the speckle patterns using the scattering of POV beams.}}
    \label{fig:expt1}
   \end{figure}
   \end{center} 

The generated POV beams have been scattered through a ground glass plate (GGP)(DG10-600, from Thorlabs) which is placed at the generation plane of POV beams, at a distance of 72.5 cm from the axicon. The resultant speckle patterns have been recorded at different propagation distances, starting with 20-70 cm at the interval of 5 cm, from GGP, and for different orders of the POV beams using the CCD camera. The experimental set up for the same has been shown in Fig. \ref{fig:expt1}. Later, we have numerically calculated the cross-correlation between two speckles patterns corresponding to the different orders of POV beam in order to generate the Bessel coherence functions.
 
\begin{center}
   \begin{figure}[htb]
   \includegraphics[width=8.0cm]{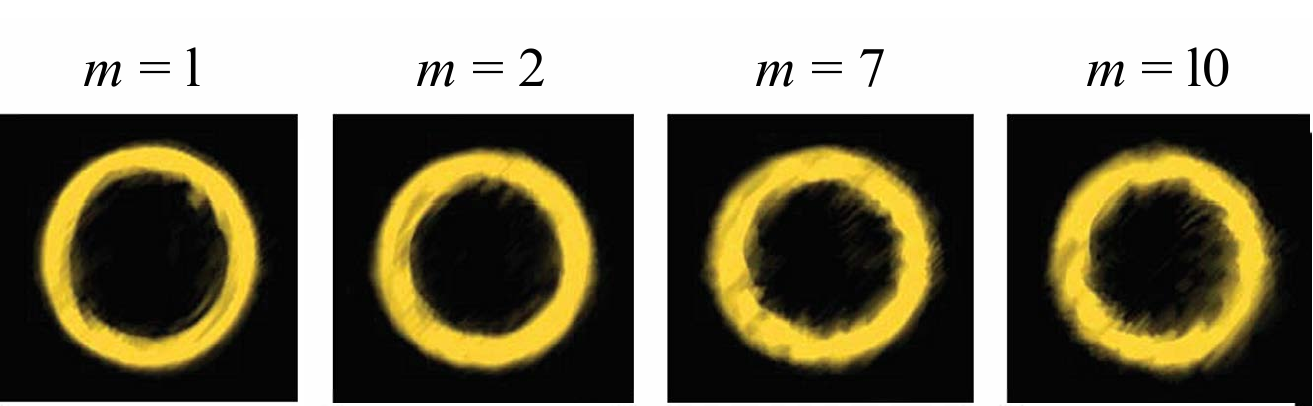}
    \caption{\textit{(Colour online) The intensity distribution of POV beams with $m$ = 1, 3, 7, 10 (from left to right) at the plane of generation.}}
    \label{fig:POV}
   \end{figure}
   \end{center}

 We start our experiment by recording the intensity distribution of the POV beams at the plane of generation. We have generated the POV beams up to the order $m$=10 using the Fourier Transform of BG beams and all orders have mode independent intensity distribution as shown in Fig. \ref{fig:POV}. These POV beams have been scattered through the GGP and the corresponding speckle patterns have been recorded through CCD.

\begin{center}
   \begin{figure}[htb]
   \includegraphics[width=8.0cm]{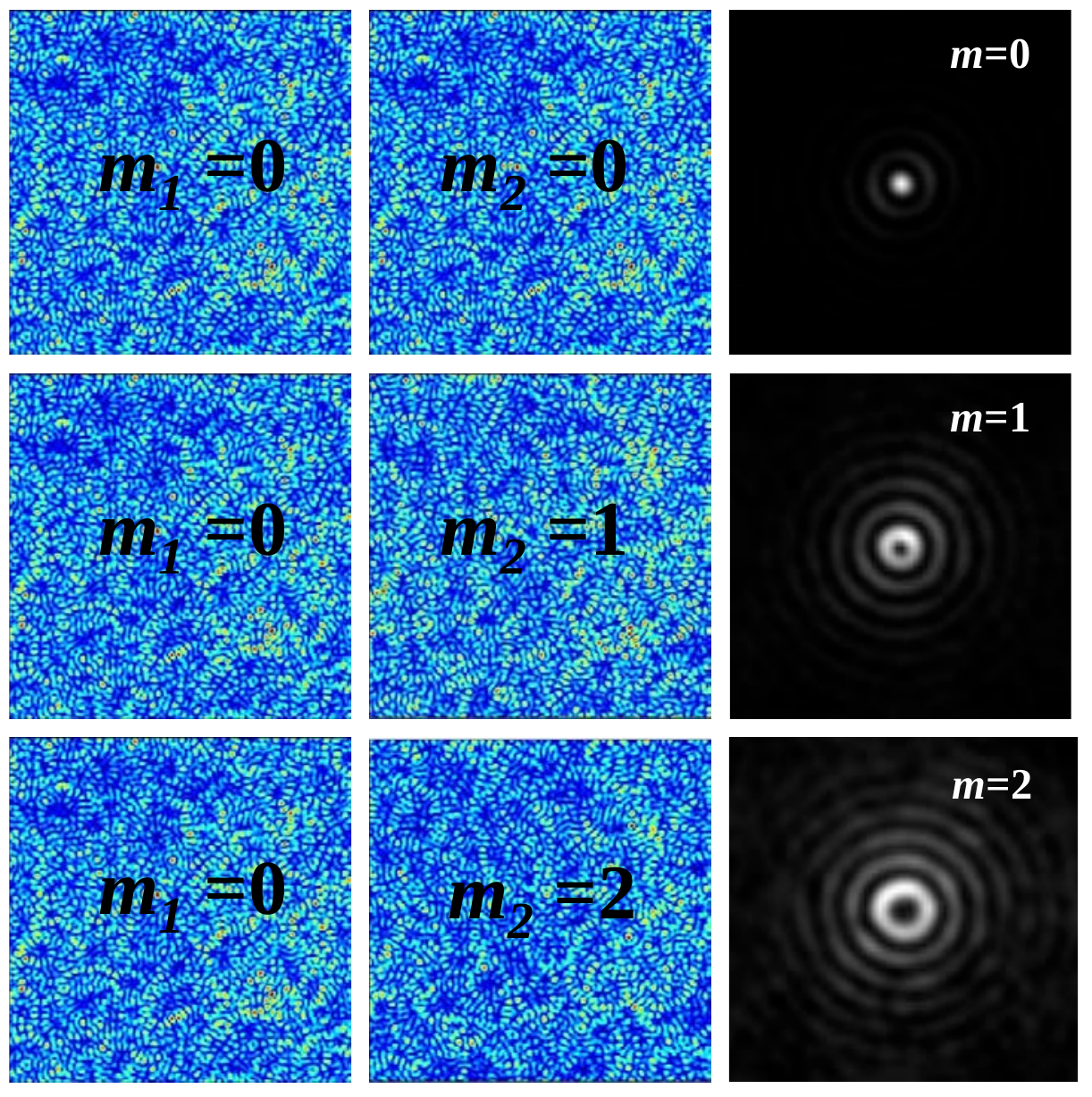}
    \caption{\textit{(Colour online) The recorded speckle patterns and the corresponding Bessel coherence functions (here $m =m_2-m_1$).}}
    \label{fig:BG}
   \end{figure}
   \end{center}   

Now, we generate the Bessel coherence functions of higher orders using cross-correlation between the speckle patterns. The order of Bessel coherence function is given by the difference between the orders of two POV beams used for producing the speckle patterns $\left( m = m_{2}-m_{1}\right)$ as represented in Eq. (7). Figure \ref{fig:BG} shows the speckle patterns as well as their cross-correlation function for different $m_{1}$ and $m_{2}$ values. The left column shows the speckle pattern corresponding to the POV beam of order $m_{1}$ = 0 and middle column shows the speckle patterns corresponding the POV beams of orders $m_{2}$ = 0-2 (from top to bottom). The last column shows the cross-correlation function between the two speckle patterns.It is clear from the figure that the auto-correlation between the speckle patterns provides the Bessel coherence function of order 0 (top row). Although we have provided the auto-correlation function for $m_{1} =0$ and $m_{2} = 0$, we have also verified the same for different orders. The cross-correlation between the speckle patterns corresponding to two different orders provides the higher order Bessel coherence function (middle and last rows). These results have been recorded at a distance of 20 cm from GGP $\left(z = 20 cm\right)$. The auto-correlation function is also consistent with the Van-Cittert-Zernike theorem which states that the auto-correlation function is given by the Fourier Transform of source plane \cite{mandel}. It is also shown that the Fourier transform of incident mode provides the information about the correlations present in speckle patterns but not the inhomogeneity size and distribution of the random medium \cite{fractal}.  Our experimental results are coinciding with the above prediction as the auto correlation function is zeroth order Bessel function which is the Fourier transform for a POV beam.

\begin{center}
   \begin{figure}[h]
   \includegraphics[width=7.0cm]{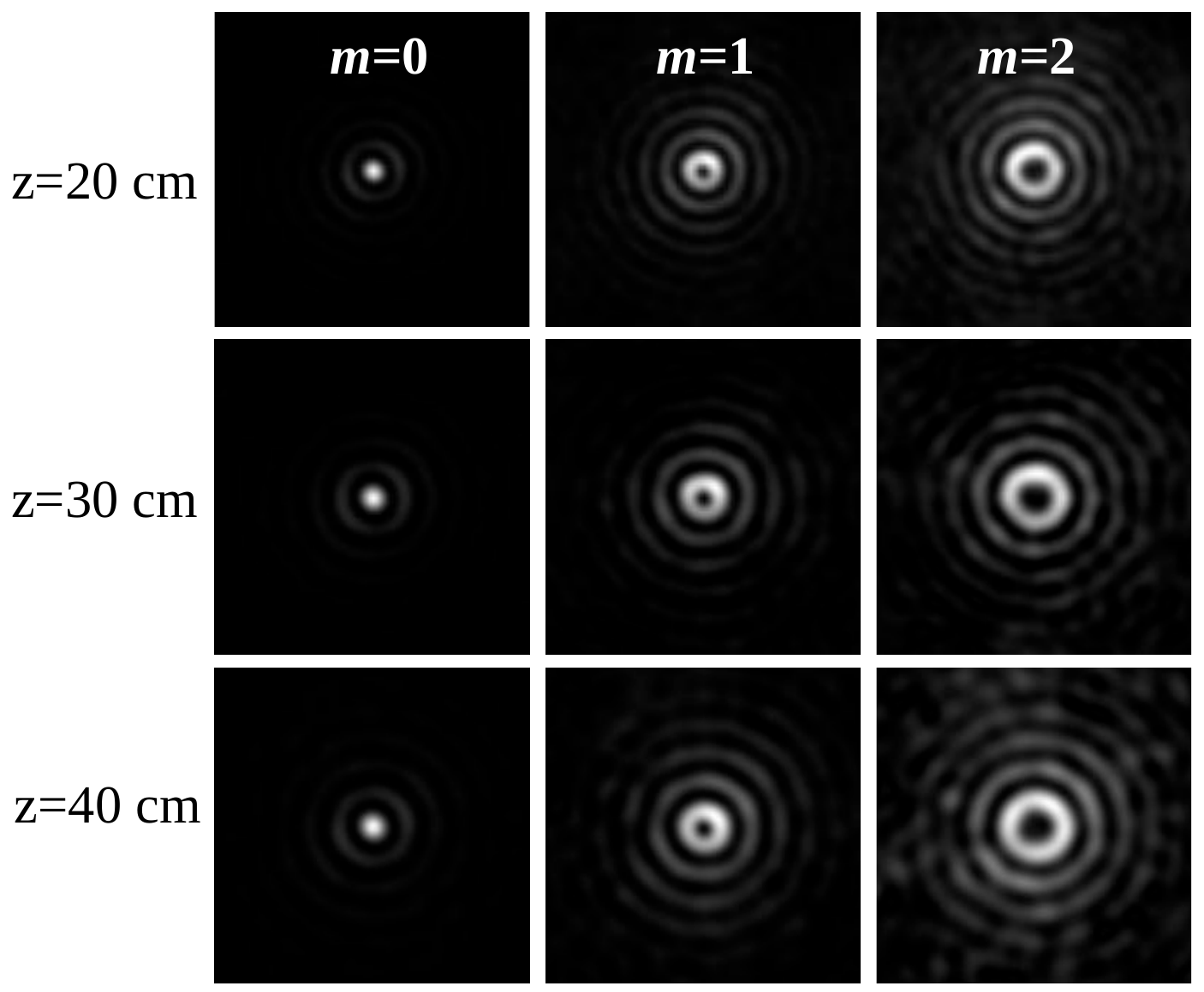}
    \caption{\textit{The experimentally generated Bessel-Gauss coherence functions with different propagation distances and with different orders $m$=0 to 2 at propagation distances $z$=20,30 and 40cm.}}
    \label{fig:BG1}
   \end{figure}
   \end{center} 
We show the Bessel coherence functions of orders $m$ = 0-2 corresponding to the different propagation distances $z$ = 20 cm, 30 cm, 40 cm in Fig. \ref{fig:BG1}. These intensity distributions have been produced through the cross-correlation between the speckle pattern corresponding to zeroth order and the speckle patterns corresponding to that particular order. These results are in agreement with the theoretical results described by the Eqs. \ref{output3}.

\begin{center}
   \begin{figure}[htb]
   \includegraphics[width=8.0cm]{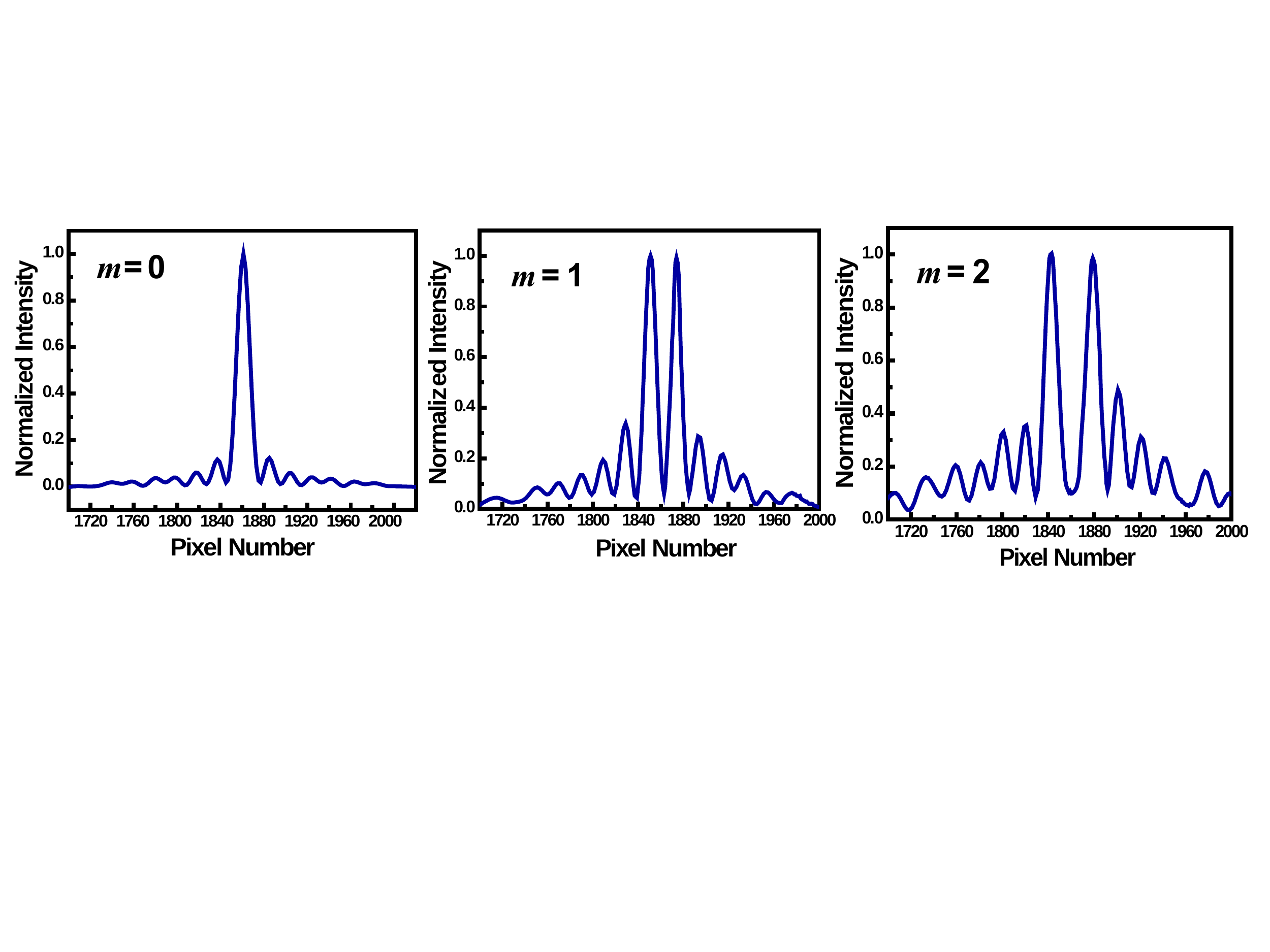}
    \caption{\textit{(Colour online) Line profiles along the center of Bessel coherence functions shown in Fig. \ref{fig:BG1} for orders $m$ = 0-2 (from left to right) at the propagation distance of $z$=40 cm.}}
    \label{fig:BG2}
   \end{figure}
   \end{center}

The Bessel coherence functions can be verified through a line profile along the center of the beams. The line profiles corresponding to the Bessel coherence functions of orders $m$ = 0-2 at a distance of $z$ = 40 cm have been shown in Fig. \ref{fig:BG2}. One can verify the presence of rings through the line profiles and confirms the Bessel nature of coherence functions. We have also determined the radius of first ring with respect to the center using these line profiles. The radius is considered as the distance between the center to peak of the first ring.

\begin{center}
   \begin{figure}[h]
   \includegraphics[width=8.0cm]{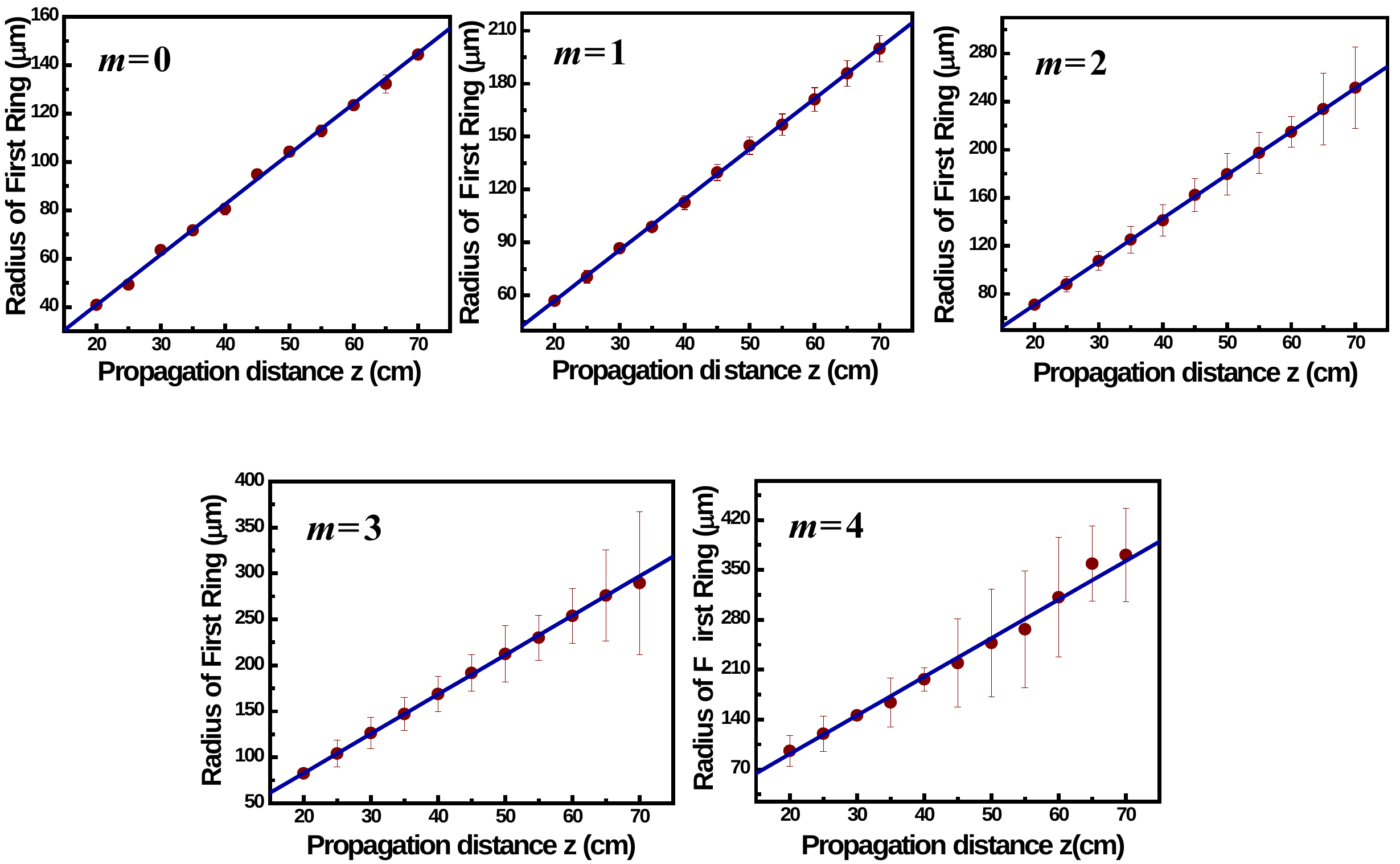}
    \caption{\textit{(Colour online) Variation of radius of the first ring with propagation distance ($z$) for Bessel Coherence functions for different orders.}}
    \label{fig:BG3}
   \end{figure}
   \end{center}

After plotting the line profiles, we have found the distance between the center to the peak of first ring for different orders $m$ = 0-4 at all propagation distances from $z$ = 20-70 cm with an interval of 5 cm. We observed that the radius of first ring varies linearly with the propagation distance for all orders from $m$ = 0-4. We have determined 14 data points for every radius using the 7 different line profiles of the beam, for different orientations. The corresponding statistical error has been shown in the graphs of Fig. \ref{fig:BG3}. For $m$ = 0 the error is nominal and it increases with the increase in the order of the Bessel coherence function as well as propagation distance. It is due to the fact that the auto-correlation function is independent of the alignment whereas the cross-correlation is alignment dependent. Even small misalignment in the camera can cause large disturbance in the cross-correlation function which we will infer as the primary source for the error at greater propagation distances as well as for higher orders.

\begin{center}
   \begin{figure}[h]
   \includegraphics[width=6.5cm]{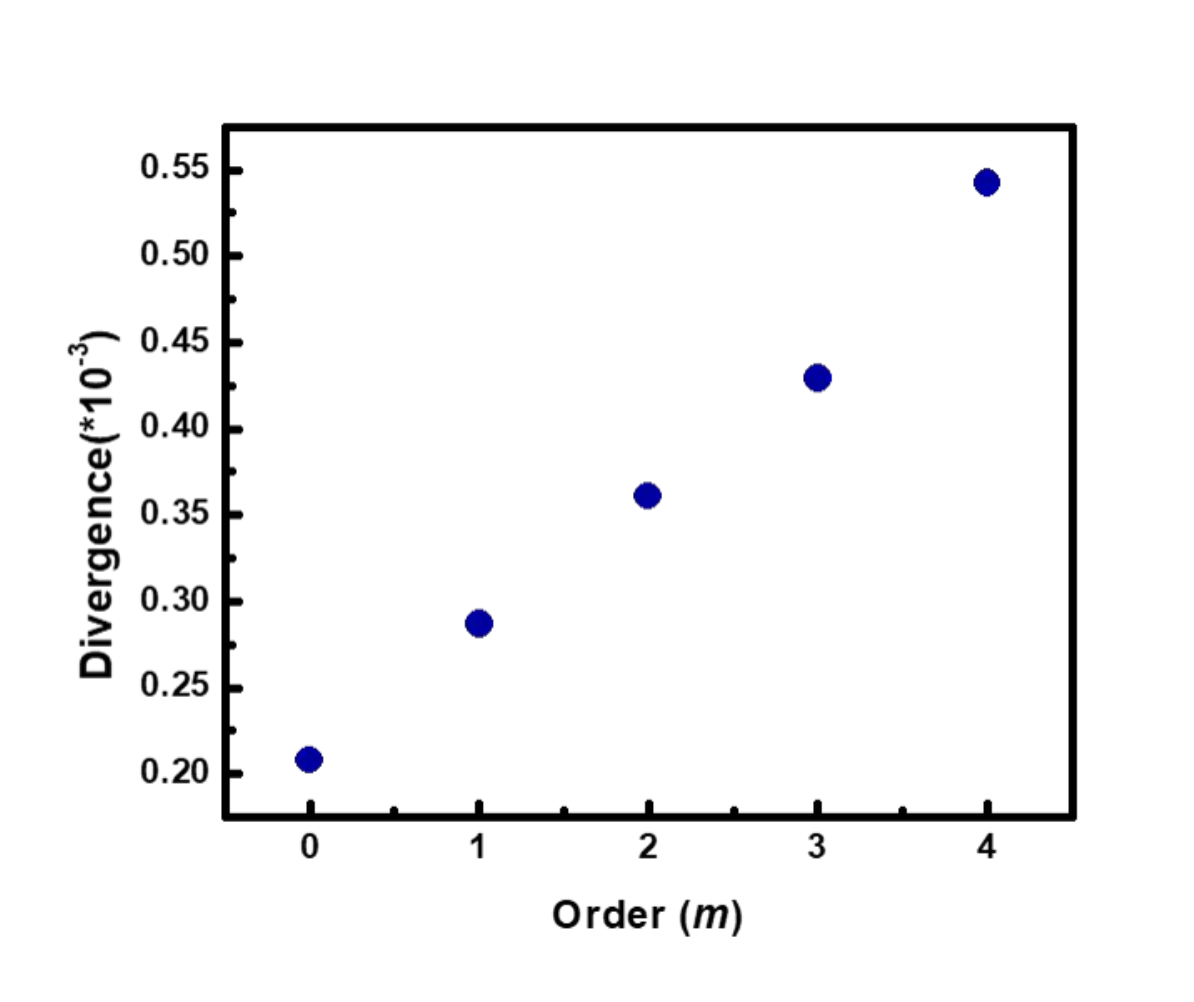}
    \caption{\textit{(Colour online) The variation of divergence with the order of the Bessel coherence functions.}}
    \label{fig:BG4}
   \end{figure}
   \end{center}

Now, we define the divergence as the rate of change of the radius of first peak with the propagation distance \cite{divergence} i.e. slope of the line which we get by using the linear fit to the data. Fig. \ref{fig:BG4} shows the variation of divergence with the increase in order for $m$ = 0-4. It is clear from the figure that the divergence varies linearly with order.

In conclusion, we have generated the Bessel coherence functions using the speckle patterns obtained by the scattering of POV beams. We have also studied their propagation and used the radius of first peak for defining divergence. We found that the radius varies linearly with propagation distance and the divergence varies linearly with order. One may use these results for generating 1-D key from a 2-D correlation image which will enhance the security over the key generated with just speckle patterns.

\end{document}